\documentclass[12pt]{iopart}
\usepackage{psfrag}

\usepackage{xspace}
\usepackage{graphics}
\usepackage{graphicx}
\usepackage{iopams}

\newcommand{\ave}[1]{\left\langle #1 \right\rangle}
\newcommand{\numave}[1]{\overline{#1}}

\newcommand{\Zset}{\mathbb{Z}}

\newcommand{\ie}{{\it i.e.}\xspace}

\newcommand{\gcomment}[1]{}

\newcommand{\slabel}[1]{\label{#1}}
\newcommand{\flabel}[1]{\label{#1}}
\newcommand{\elabel}[1]{\label{#1}}

\newcommand{\WP}{\widetilde{\psi}}
\newcommand{\OC}{{\mathcal{O}}}
\newcommand{\erf}{{\mathcal{E}}}
\newcommand{\half}{\frac{1}{2}}

\begin{document}
\title[Multiscaling in areas between random walkers]
{Multiscaling in the sequence of areas enclosed by coalescing random walkers}
\author{Peter Welinder\dag, Gunnar Pruessner\dag\ddag\ and Kim
Christensen\dag}

\address{\dag\ Condensed Matter Theory Group,
Blackett Laboratory,
Imperial College London,
Prince Consort Rd,
London SW7~2BW,
UK}
\address{\ddag\ Mathematics Institute,
University of Warwick,
Gibbet Hill Road,
Coventry CV4 7AL,
UK}
\date{\today}

\ead{gunnar.pruessner@physics.org}

\begin{abstract}
We address the question whether the sequence of areas between
coalescing random walkers displays multiscaling and in the process calculate
the second moment as well as the two point correlation function exactly. The
scaling of higher order correlation functions is estimated numerically, indicating
a logarithmic dependence on the system size. 
Together with the analytical results, this confirms the presence of
multiscaling.
\end{abstract}

\submitto{NJP}


\maketitle

\section{Introduction}
Gap scaling is found frequently in the context of scale invariance, such
as equilibrium statistical mechanics of phase transitions
\cite{PfeutyToulouse:1977},
growth phenomena \cite{Krug:1997}, reaction diffusion
processes \cite{benAvrahamHavlin:2000}
or self organised criticality
\cite{Jensen:1998,ChristensenMoloney:2005}. In the
present context, gap scaling \cite{PfeutyToulouse:1977}
means that the $n$th moment of the observable $s$, denoted as
$\ave{s^n}$, in leading order scales like $t^{\Delta(1+n-\tau)}$ in some
large parameter $t$. The ``gap'' refers to the constant difference
$\Delta$ between two exponents for $n$ and $n+1$.
In a stochastic process recently introduced as the ``totally asymmetric
Oslo model'' \cite{JensenPruessner:2003,Pruessner_exactTAOM:2003}, the moments of the area
under a random walk along an absorbing wall was found to obey gap
scaling with $\tau=4/3$ and $\Delta=3/2$. 

If the moments scale asymptotically as power laws (possibly with
logarithmic corrections) but the exponents do not increase
linearly with $n$, they are said to display multiscaling.
The presence of multiscaling is sometimes confused with the absence of
scaling altogether and very few examples of simple dynamical processes
are known, which can be shown by exact calculation to display
multiscaling. It is therefore
highly desirable to find a simple stochastic process, such as Brownian
motion, for which multiscaling can be derived analytically. Motivated by a
recent study of cluster aggregation
\cite{ConnaughtonRajeshZaboronski:2005,ConnaughtonRajeshZaboronski:2006}, we
study the scaling of ``sequential moments'' or, more accurately,
correlation functions of the area size between the
trajectories of
coalescing random walkers, illustrated in \Fref{trajectories_x01}. 
We hope that future research will make the link between our results and
the results recently obtained by Munasinghe \etal \cite{MunasingheETAL:2006b} on the
scaling of the $n$-point correlation
function
of coalescing random walkers.

Recent years have seen a considerable interest in the distribution of the
area size under a random walker trajectory. It has been investigated in
the context of ``directed rice piles''
\cite{Pruessner_exactTAOM:2003,StapletonChristensen:2006}, as
well as in the context of extreme value statistics
\cite{MajumdarComtet:2004,MajumdarComtet:2005}.  In both cases, the
object of interest is the area between trajectories of coalescing
random walkers with initial spacing $x_0$, or, equivalently, the area
under the trajectory of a random walker,
which starts at time $t'=0$ at distance $r(t'=0)=x_0$ away from an absorbing
wall. In the next section, the model will be introduced in detail.

Addressing problems of extreme value statistics, Majumdar and Comtet
\cite{MajumdarComtet:2005} constrain the ensemble to trajectories with
$r(t'=0)=x_0$ and, in addition, with $r(t'=t)=x_0$ at a particular
``termination time'' $t'=t$ in the limit $x_0\to0$. In this limit, they
derive very elegantly the exact distribution function of the area
sizes.  Due to the lack of additional scales, and, correspondingly, due to
the lack of a dimensionless parameter, \emph{in the case $x_0\to0$ the
\emph{scaling} of all observables is readily derived from dimensional
analysis}, \ie the exponents are immediately known and gap-scaling is a
necessity for all existing moments. 
In contrast, we will be concerned with finite $x_0$ (so that \emph{a
priori} neither exponents nor type of scaling are known) 
and an ensemble not
constrained by the termination time in the form $r(t'=t)=x_0$.
This
problem has been addressed earlier \cite{StapletonChristensen:2006} using
some of the results in \cite{MajumdarComtet:2005} for a calculation of
the distribution function of the areas in leading order of $x_0$.

Unfortunately, only the first moment, which is trivial, is known
exactly. Very little is known about higher moments beyond leading order.
We employ analytical calculations, based on the formalism
introduced in \cite{Pruessner_exactTAOM:2003} as well as extensive
numerics to overcome this problem. In the course, we determine the exact
second moment of the distribution, as well as the two point correlation
function, which displays anti-correlations, sometimes interpreted as a hallmark of
multiscaling \cite{MunasingheETAL:2006b,MunasingheETAL:2006}.
To our knowledge, these are the first non-trivial exact
results for the area size distribution under a random walker with
$x_0>0$.

At least three correlation functions are required to decide about the presence of
multiscaling.  So
far, we have been unable to perform the calculation of the third moment  
analytically. We
have therefore resorted to Monte Carlo techniques, which are,
unfortunately, hampered by slow convergence due to rare events.
Nevertheless, the numerics strongly suggests multiscaling.

In the next section, we will introduce the model and its observables.
The third section contains most of the analytical and
numerical results, which are briefly summarised in the last section.

\section{Model}
\slabel{model}
\Fref{trajectories_x01} shows a lattice realisation (see below) of a sequence of random
walkers $i=0,1,2,\ldots$ with diffusion constant $D_0$ and trajectories
$r_i(t')$ which start at $t'=0$
from $r_i(t'=0)=i x_0$. Whenever two random walkers meet, they
coalesce
and the resulting, single walker continues its walk. The walkers are
indistinguishable and it is therefore irrelevant whether walkers
actually run simultaneously or sequentially, in which latter case they are thought
to stop as soon as they intersect the trajectory of another walker. This
is the picture we will adopt henceforth.

\begin{figure}
\begin{center}
\includegraphics[scale=0.7]{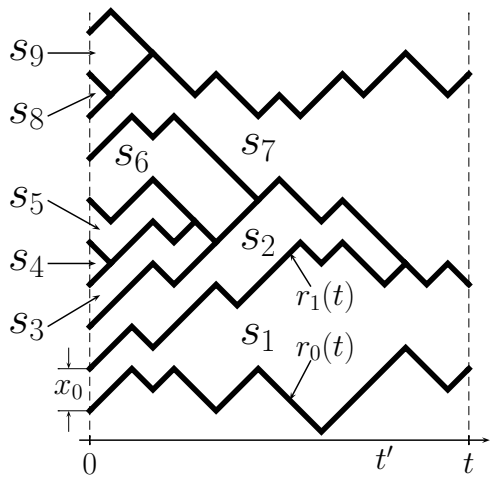}
\end{center}
\caption{\flabel{trajectories_x01}
A sequence of trajectories $r_0(t'), r_1(t'),\ldots$ 
starting at $t'=0$ with initial spacing $x_0$ and demarcating areas
$s_1,s_2,\ldots$ between them. The dashed lines indicate the ``walls''
at $t'=0$ and the terminating time $t'=t$.
Whenever random walkers meet, they coalesce
and perform the rest of the walk together. 
}
\end{figure}

For the sake of definiteness, all trajectories are considered to run
from $t'=0$ to $t'=t$. If two walker coalesce, their trajectories
coincide for the remaining time. The area $s_i$ between two walkers $i-1$ and
$i$ with
trajectories $r_{i-1}(t')$ and $r_i(t')$ respectively,
\begin{equation}
s_i = \int_0^t \rmd t' \left(r_i(t') - r_{i-1}(t')\right) \ge 0
\end{equation}
is a random variable, the moments of which will be denoted by
$\ave{s^n}$, where $\ave{ }$ is the expectation value over all
configurations.  

Given an individual trajectory, it is impossible to
tell whether it is produced by merging walkers or not. Producing the
trajectories sequentially, the only trajectory relevant to the fate of
walker $i$ is the trajectory $i-1$. There is no transient (other than
producing the very first trajectory), \ie stationarity is reached
immediately. That means, for example, that the expectation value
$\ave{s_1 s_2}$ is translation invariant, $\ave{s_1 s_2}=\ave{s_i
s_{i+1}}$ for all $i\ge1$.

If the random walkers are unbiased, any sequence of areas
$(s_i,s_{i+1},\ldots,s_{i+n})$ is, by construction, as likely as the
sequence $(s_{i+n},s_{i+n-1},\ldots,s_i)$. This can be seen by mirroring
the entire sequence which, in case of an unbiased walk, leaves the
probabilities of individual trajectories unchanged.  
In the presence of a bias 
this is no longer
case and so the joint probability of a
sequence changes under inversion.  Thus, in the presence of a bias, one
expects, for example, $\ave{s_1^2 s_2}\ne\ave{s_2^2 s_1}$, while the two
averages would be equal in the absence of a bias. Of course, the observable
itself might be inversion invariant, for example $\ave{s_1 s_2}=\ave{s_2
s_1}$ regardless of a bias.

As long as the two paths $r_i$ and $r_{i-1}$ have not merged, the
variance of the difference $r_i-r_{i-1}$ corresponds to the variance of
a random walker with twice the diffusion constant $D=2D_0$, walking
along an absorbing wall.  However, while the individual trajectories $r_i$ might be biased, \ie
in the presence of a drift, the difference does not have such a bias.
It is therefore clear that whatever is said about the (single) area size
distribution between random walkers with initial spacing $x_0$, biased
or not, equally applies to the (single) area size distribution under an unbiased
random walker with twice the diffusion constant starting at distance
$r(t=0)=x_0$ from an absorbing wall. However, the notion of a sequence
is missing in the latter case.

The main aim of this work is to calculate 
the correlation functions $\ave{s_1 s_2}$, $\ave{s_1 s_2 s_3}$, and $\ave{s_1 s_2
s_3 s_4}$ and determine their asymptotes, that means the exponent
$\gamma_n$ and possibly the polynomial $\mathcal{P}$ so that
\begin{equation}
\lim_{t\to\infty}\frac{\ave{s_1 s_2\ldots s_n}}{t^{\gamma_n} \mathcal{P}(\ln t) } = c_n >0
\end{equation}
suspecting that $\gamma_n$ is not linear in $n$ and/or that
$\mathcal{P}$ is non-trivial.

As it turns out, $\ave{s_1 s_2}$ and indeed $\ave{s_1
s_{n+1}}$ can be derived from $\ave{s^2}$, the calculation of which is
detailed in the first part of the result section.
The other two moments could only be tackled numerically, presented
in the second part of the result section.

Most asymptotes will be determined in the limit of large $t$. 
It is important to realise the r{\^o}le $x_0$ plays in this context. The
problem is fully parametrised by $t$, $D$ and $x_0$.
With the
area having the same dimension as $t \sqrt{D t}$, 
in case of vanishing $x_0$
the scaling of
the expectation value of \emph{any} product of these areas, say
$\ave{s_1 s_3^2}$, is fixed to be a corresponding  power of $t \sqrt{D
t}$, \ie $\ave{s_1 s_3^2}\propto t^{9/2}$. In fact, there cannot even be
any corrections to this behaviour. However, if $x_0\ne0$, there is a
dimensionless quantity $Dt/x_0^2$, which can and, indeed, does change
the exponents of the asymptotes and gives rise to corrections to scaling
\cite{Wegner:72}.

\section{Results}
In this section, we will first discuss in detail the analytical results,
starting with a brief review of the method which is used to
calculate the second moment $\ave{s^2}$ in closed form. After deriving its expansion in powers of
$t$, some shortcomings of the perturbative approach of
\cite{Pruessner_exactTAOM:2003} are discussed. Using these results, the
correlation function $\ave{s_1 s_{n+1}}$ is derived, which includes the
original problem $\ave{s_1 s_2}$ as a special case. This part finishes
with a discussion on the effect of anti-correlations and a brief section
on higher moments.

The second part of the result section is concerned with our numerical
results. First, the lattice effects are estimated by a comparison of the
numerical estimate for $\ave{s_1 s_2}$ with the analytical result. After
discussing some technical problems, the results for $\ave{s_1 s_2 s_3}$
and $\ave{s_1 s_2 s_3 s_4}$ are shown, indicating the presence of
logarithmic corrections.

\subsection{Exact calculations}
When calculating 
correlation functions
of the form $\ave{s_1 s_2}$, one
of the central observations is that they can be obtained by considering
``local moments'' of the form $\ave{s^2}$ in a system with twice the
initial spacing, 
\footnote{We thank Alan Bray for sharing this insight with
us.} which will be discussed in detail below. 
We therefore start the
section with the calculation of the second moment and its
implications for earlier results, in particular the approximation scheme
used in \cite{Pruessner_exactTAOM:2003}. Knowing $\ave{s^2}$ then allows us
to calculate $\ave{s_1 s_{n+1}}$.

All analytical calculations are based on the hierarchy of differential
equations
introduced in \cite{Pruessner_exactTAOM:2003}, 
\begin{equation} \elabel{pde}
  \partial_t \psi_n(t,x;x_0) = D \partial_x^2 \psi_n(t,x;x_0) + x n \psi_{n-1}(t,x;x_0)
\end{equation}
with boundary conditions
\begin{eqnarray}
 \lim_{t\to 0} \psi_n(t,x;x_0) & = & \delta_{n,0} \delta(x-x_0)
 \elabel{bc_space} \\
 \psi_n(t,0;x_0) & = & 0 \elabel{absorbing}
\end{eqnarray}
Starting with $n=0$, $\psi_0(x,t; x_0)$ is the probability to find at
$x$ a fair random
walker along an absorbing wall
with diffusion constant $D=2D_0$ that started at $x_0$. For $n>0$,
$\psi_n$ is the local contribution to the $n$-th moment of the area
under the trajectory, see \cite{Pruessner_exactTAOM:2003}. For $n=0$ one
finds, consistently, $\ave{s^0}=1$, for $n>0$ one has
\begin{equation}
 \ave{s^n}(t;x_0) = \int_0^t \rmd t' \int_0^\infty \rmd x' x' n
 \psi_{n-1}(t',x';x_0)  \ .
 \elabel{def_sn}
\end{equation}
To ease notation, the dimensionless form of $\psi_n$ is introduced as
\begin{equation} \elabel{def_wp}
\psi_n(x,t;x_0) = \frac{1}{x_0} \left(\frac{x_0^3}{D}\right)^n \WP_n(y,\tau) 
\end{equation}
with
$y=x/x_0$ and $\tau=t/(x_0^2/D)$, so that the differential equation
reads
\begin{equation}\elabel{dimless_pde}
  \partial_\tau \WP_n(\tau,y) = \partial_y^2 \WP_n(\tau,y) + y n
  \WP_{n-1}(\tau,y)
\end{equation}
with $\delta(x-x_0)$ in boundary condition \eref{bc_space} replaced by
$\delta(y-1)$. The propagator $G(y,\tau;y_0)$ of
\Eref{pde} can be constructed easily,
\begin{equation}
 G(y,\tau;y_0) \equiv \frac{1}{\sqrt{4 \tau \pi}} \left( 
e^{-\frac{(y-y_0)^2}{4\tau}} - e^{-\frac{(y+y_0)^2}{4\tau}}
\right) \ ,
\end{equation}
which at $y_0=1$ coincides with $\WP_0(y,\tau)$, i.e.
\begin{equation}
 \WP_0(y,\tau)=G(y,\tau;1)= \frac{1}{\sqrt{\tau \pi}} e^{-\frac{y^2+1}{4 \tau}} \sinh\left(\frac{y}{2\tau}\right)
\ .
\elabel{def_WP0}
\end{equation}
The formal solution is the hierarchy
\begin{equation}
 \WP_n(y,\tau) = \int_0^\tau \rmd \tau' \int_0^\infty \rmd y' n y' \WP_{n-1}(y',\tau') G(y,\tau-\tau';y') 
\elabel{hierarchy}
\end{equation}
which is, as it turns out, not easily calculated in closed form. Only
the first moment is calculated immediately 
\begin{equation}
\ave{s}(t;x_0) = \int_0^t \rmd t' \int_0^\infty \rmd x' x' 
 \psi_{0}(t',x';x_0) = x_0 t \ .
 \elabel{first_moment}
\end{equation}

It has been shown \cite{Pruessner_exactTAOM:2003,StapletonChristensen:2006} that the moments
follow gap scaling,
\begin{equation}
\lim_{t\to\infty} \frac{\ave{s^n}(t;x_0)}{t^{\Delta(1+n-\tau)}} = x_0 
D^{(n-1)/2} C_n
\end{equation}
with $\tau=4/3$ and $\Delta=3/2$ and $C_n$ dimensionless.

Using the hierarchy \eref{hierarchy} and the definition of the 
moments \eref{def_sn}, the second moment is
\begin{eqnarray}
\fl
 \ave{s^2}(t;x_0) & = &
 2 \left(\frac{x_0^3}{D}\right)^2
 \int_0^T \rmd \tau  \int_0^\infty \rmd y  y 
 \int_0^\tau                       \rmd \tau' \int_0^\infty \rmd y' y' \nonumber \\
 && \lo\times
 \frac{1}{\sqrt{4(\tau-\tau')\pi}} 
   \left(e^{-\frac{(y-y')^2}{4(\tau-\tau')}} - e^{-\frac{(y+y')^2}{4(\tau-\tau')}}\right)
 \frac{1}{\sqrt{4\tau'\pi}} 
   \left(e^{-\frac{(y'-1)^2}{4\tau'}} - e^{-\frac{(y'+1)^2}{4\tau'}}\right)
   \elabel{s2_integral}
\end{eqnarray}
where $T=T(t;x_0)=\frac{t}{x_0^2/D}$. The integration of \eref{s2_integral}
would be fairly straight forward, was it not for the factors $y$
and $y'$. 

There is no reason to expect convergence problems from the improper
integrals, because the Gaussians (and even more so differences of
Gaussians)
effectively cut them off. We therefore feel confident when
changing the integration order, so that the improper integrals are done
first and in reverse order. The integration over $y$ can be done
immediately and produces only a factor of the form
$y'\sqrt{4(\tau-\tau')\pi}$. The integral over $y'$ is slightly more
complicated, as it is of the form $\int \rmd y' y'^2 G(y',\tau';1)$, and it
produces an exponential and an error function, all multiplied by powers
of $\tau'$. Remarkably, the resulting integrand for the integration over
$\tau'$ and $\tau$ is independent of $\tau$, so that the integral over
$\tau'$ can be replaced by the same expression evaluated at $\tau$ 
with a pre-factor $T-\tau$. Due to the presence of
pre-factors involving various powers of $\tau$, the final integration over
$\tau$ produces many different terms:
\begin{eqnarray} \fl
  \ave{s^2} (t; x_0) & \lo= & \frac{1}{180} \left( \frac{x_0^3}{D} \right)^2
  \Bigg[ 
      (1 + 28 T + 132 T^2)  \frac{2}{\sqrt{\pi}} \sqrt{T} e^{-\frac{1}{4 T}}
    - (1 + 30 T) \nonumber \\
    && \lo+ (1 + 30 T + 180 T^2 + 120 T^3 ) \erf \left( \frac{1}{\sqrt{4 T}} \right)
    \Bigg]  
    \elabel{s2_exact} \\
       &\lo=&  \frac{1}{180} \left( \frac{x_0^3}{D} \right)^2  
  \left[ 
       \left(1 + 28 \frac{t D}{x_0^2} + 132 \left( \frac{t D}{x_0^2} \right)^2 \right) 
        \frac{2}{\sqrt{\pi}} \frac{\sqrt{t D}}{x_0}  e^{-\frac{x_0^2}{4 t D}} 
     - \left( 1 + 30 \frac{t D}{x_0^2} \right)  \right.
     \nonumber \\
    && \lo+
     \left.
     \left(  1 
           + 30 \frac{t D}{x_0^2} 
           + 180 \left(\frac{t D}{x_0^2}\right)^2
           + 120 \left(\frac{t D}{x_0^2}\right)^3
     \right)
       \erf \left( \frac{x_0}{\sqrt{4 t D}} \right) \right] 
\end{eqnarray}
where $\erf$ is the error function $\erf(u)=(2/\sqrt{\pi})\int_0^u \rmd x e^{-x^2}$.
To our knowledge this is the first non-trivial moment of the area
distribution under a random walker that has been calculated exactly.

It is very instructive to expand the result in powers of
$T=T(t;x_0)=t/(x_0^2/D)$:
    \begin{eqnarray}\fl
      \frac{\ave{s^2}(t; x_0)}{(x_0^3/D)^2} &\lo=&
        \frac{32}{15 \sqrt{\pi}} T^{5/2} 
      + \frac{8}{9 \sqrt{\pi}} T^{3/2} 
      - \frac{1}{6} T 
      + \frac{1}{15 \sqrt{\pi}} T^{1/2} 
      - \frac{1}{180}
      + \frac{1}{1260 \sqrt{\pi}} T^{-1/2} \nonumber \\
      &&\lo- \frac{1}{181440 \sqrt{\pi}} T^{-3/2}
      + \frac{1}{13305600 \sqrt{\pi}} T^{-5/2}
      + \mathcal{O} \left( T^{-7/2} \right).
    \elabel{s^2_expansion}
\end{eqnarray}
with the terms beyond leading order to be considered corrections to
scaling \cite{Wegner:72}.
There are two terms without a factor $\pi^{-1/2}$, namely
$- \frac{1}{6} T$ and
$- \frac{1}{180}$,
which are present in the same form, \ie without any further factor, already in
the exact expression \eref{s2_exact}.
What
makes these terms very different from all others is that they are not
an odd-half power of $T$. In fact, to any order, \emph{all} terms apart
from $-(1+30 T)/180$, are odd-half powers of $T$. In
\cite{Pruessner_exactTAOM:2003} only the odd-half powers had been anticipated,
suggesting that no deviation from this form is possible, 
but that
turns out to be incorrect now.

It is instructive to understand why the systematic, perturbative expansion in
\cite{Pruessner_exactTAOM:2003} misses the two terms. It relies on a perturbative expansion of 
$\WP_0(\sqrt{\mu} y,\mu \tau)$, \eref{def_WP0}, for large $\mu$ at fixed, finite $y$ 
and $\tau$. The scaling in $\mu$ is then inherited by $\WP_1(\sqrt{\mu}
y,\mu \tau)$ through the integral \eref{hierarchy} and by $\ave{s^2}(\mu
t; x_0)$ through \eref{def_sn}. This
relation gives rise to the scaling of $\ave{s^2}(t;x_0)$ by
parameterising $t$ in convenient units, $t=\mu t_0$.
Along the lines of a renormalisation group
calculation \cite{Taeuber:2005}, 
$\mu$ can be considered the ``flow parameter'' and
$t_0=x_0^2/D$ as the normalisation point. Using this procedure for all
orders in $\mu$ produces the prediction of only odd-half powers of
$\mu$.

However,
when evaluating the perturbative expansion of $\WP_0(\sqrt{\mu} y,\mu \tau)$ in powers of $\mu$ in the
integrals producing $\WP_1$ and $\ave{s^2}$, one violates the
premise that $y$ and $\tau$ are finite, as in the integral
\eref{hierarchy} $\tau$ runs from $0$ to a finite value and the integral
over $y$ is to be evaluated at divergent upper bound. As a result, one
finds integrals of the form 
$\int_0^T\rmd \tau\int_0^{\tau}\rmd \tau' \tau'^{1/2-i}$ 
for $i=0,1,2,\dots$. For $i=0,1$ the integrals are
convergent, producing terms of
order $T^{5/2}$ and
$T^{3/2}$ respectively, however for
$i\ge2$ one ($i=2$) or both ($i>2$) integrals diverge in the lower limit. 
Ignoring
this divergence, \ie only keeping the upper limit, 
produces the correct amplitudes nevertheless. 
We have tested the integrals for
$i=0,1,\dots,5$ by direct evaluation 
and comparison to the exact result in the form of the Taylor expansion
\eref{s^2_expansion}, of course, with the terms of order $T^1$ and $T^0$
missing. In general, the integrals for the perturbative expansion give the following odd-half powers:
\begin{eqnarray}
\ave{s^2}(T;x_0) 
+
\left( \frac{x_0^3}{D} \right)^2
\frac{1+30T}{180}
\nonumber \\
= \half \frac{1}{\sqrt{\pi}} 
\left( \frac{x_0^3}{D} \right)^2
\sum_{i=0}^\infty
T^{5/2-i}
\sum_{j=0}^i
\frac{4^{2+j-i}}{\left(\frac{3}{2}-i\right)\left(\frac{5}{2}-i\right)}
\frac{(j+1)! (-)^{i-j}}{(2j+1)! (i-j)!} \ .
\end{eqnarray}
The terms $1/(3/2-i)$ and $1/(5/2-i)$ are reminiscent of the
integration 
$\int_0^T\rmd \tau\int_0^{\tau}\rmd \tau' \tau'^{1/2-i}$.
As long as none of the coefficients or the error term is divergent, the
series expansion can be considered a simple Taylor series in $\mu$ and therefore in particular
the leading order terms are reliable, as in an asymptotic expansion.

Similarly, 
one can use the perturbative expansion scheme to determine the amplitude
of higher orders of $\ave{s^3}$. We can confirm agreement with
\cite{StapletonChristensen:2006} of the amplitude of the leading order $T^4$, 
expecting the next to leading
order $T^3$ to be correctly predicted in the perturbative approach 
as well. 
Following the integration for
$\ave{s^2}$, now three integrals over $\tau$ might diverge and one might speculate that the first
term missed in the perturbative expansion is $T^{5/2}$ and the (actually
divergent) amplitude for $T^2$ to be correct, a $T^{3/2}$ to be missed,
$T^1$ to be correct, $T^{1/2}$ missed again, and terms of order $T^0$ and
lower to be correct again.

We now turn back to the expansion of the exact result, \Eref{s2_exact}.
When it comes to the large distance behaviour of the correlation
function, the second moment $\ave{s^2}(t;x_0)$ will be evaluated for
\emph{small} $T$. Expanding \Eref{s2_exact} for small $T$ gives
\begin{eqnarray}\fl
\ave{s^2}(t;x_0) 
\elabel{s2_expansion_small_T}
\\
\fl
\qquad=
\left(\frac{x_0^3}{D}\right)^2
\left[ T^2 + \frac{2}{3} T^3 +
e^{-\frac{1}{4T}} \frac{1}{\sqrt{\pi}}
\left(
-512 T^{13/2} + 28672 T^{15/2} - \OC\left(T^{17/2}\right)
\right)
\right] \ .
\nonumber
\end{eqnarray}
Again, the leading, polynomial orders, which together with the pre-factor
give just $\ave{s^2}(t;x_0)=x_0^2 t^2 + (2/3) Dt^3+\ldots$, 
are \emph{solely} due to the
two unusual terms with integer powers identified in \eref{s2_exact}.
However, it
is consistent with the method in \cite{Pruessner_exactTAOM:2003} being a
large $T$ expansion, that they would have been missed there. All other
terms in \eref{s2_expansion_small_T} are exponentially suppressed with a
remarkably high leading power in the polynomial. In fact, the first
\emph{six} terms, starting with $T^{1/2}$, all turn out to have vanishing
amplitude.

\subsubsection{Two point correlation function}
Based on $\ave{s^2}$ one can 
derive the two point correlation function by noting that 
\begin{equation}
\ave{(s_1 + s_2)^2} - \ave{s_1^2} - \ave{s_2^2} = 2 \ave{s_1 s_2}
\end{equation}
and more generally with $S=s_2+s_3+\ldots+s_{n-2}$ for $n\ge3$ (where
$S=0$ at $n=3$)
\begin{eqnarray}\fl
    \ave{(s_1+S)^2}(t; x_0)
- 2 \ave{(s_1+S+s_{n-1})^2}(t; x_0) 
+   \ave{(s_1+S+s_{n-1}+s_n)^2}(t; x_0) &&\nonumber \\
\lo= 2 \ave{s_1 s_n}(t; x_0) \equiv 2 c_{n-1}(t; x_0) &&
\elabel{def_cn}
\end{eqnarray}
using translational invariance in the form $\ave{s_n^2}=\ave{s_{n-1}^2}$ 
and $\ave{(s_1+S)s_{n-1}}=\ave{(S+s_{n-1})s_n}$. With
$u_n(t; x_0)=\ave{(s_1+\ldots+s_n)^2}(t; x_0)$ and the convention $u_{-1}=u_1$ and $u_0=0$, 
the correlation function is
just the lattice Laplacian of the second moment $u_n(t; x_0)$,
\begin{equation}
2 \ave{s_1 s_{n+1}}(t; x_0) = u_{n+1}(t; x_0)-2u_n(t; x_0)+u_{n-1}(t; x_0) = \nabla_n^2 u_n
\elabel{cn_laplacian}
\end{equation}
for $n\ge0$. Given \Eref{s2_exact}, the second moment $u_n(t; x_0)$ is
easily
calculated because $u_n(t; x_0)=u_1(t; n x_0)=\ave{s^2} (t; n x_0)$.
This is illustrated in \Fref{trajectories_merged}:
The network of trajectories is pruned 
by joining $n$ consecutive areas
$s_1,\ldots,s_n$ to one large area. Thereby a sequence
of larger areas is constructed, as if produced by random walkers with 
initial spacing $n x_0$. The exact two point correlation function therefore
is
\begin{eqnarray}\fl
\ave{s_1 s_{n+1}}(t; x_0) & = & 
\half
\Big(
\ave{s^2}(t; |n-1|x_0)
- 2 \ave{s^2}(t; |n| x_0)
+ \ave{s^2}(t; |n+1| x_0)
\Big)
\elabel{lattice_Laplacian}
\end{eqnarray}
for $n\in\Zset$,
with the exact expression given in \eref{s2_exact}. The modulus
guarantees the validity even for $n\le0$, although the above results
for $\ave{s^2}(t;x_0)$ have been
derived using $x_0>0$.

\begin{figure}
\begin{center}
\includegraphics[scale=0.7]{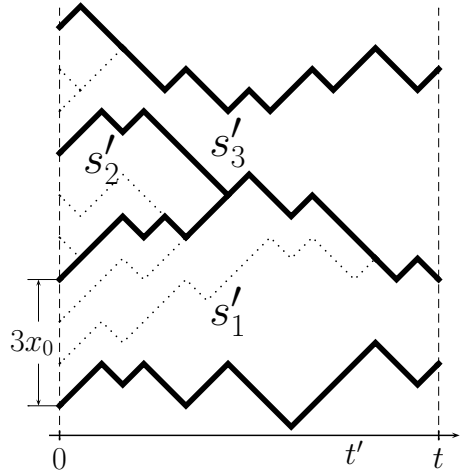}
\end{center}
\caption{\flabel{trajectories_merged}
Pruning the
network of trajectories, such as the one shown in
\fref{trajectories_x01}, by leaving only every $n$th trajectory 
produces an ensemble
of walkers with initial spacing $n x_0$ (trajectories removed are shown
dotted). Pruning is identical to merging the enclosed
areas, in this example $n=3$ so that 
$s_1'=s_1+s_2+s_3$, $s_2'=s_4+s_5+s_6$,\ldots.
Such a joining/pruning scheme, mapping an ensemble
with initial spacing $x_0$ onto an ensemble with initial spacing $n x_0$ is exact, 
even on the
lattice.
}
\end{figure}

In the following, the asymptotes of $\ave{s_1 s_{n+1}}$ are
examined.
We first consider the behaviour of $\ave{s_1 s_{n+1}}(t; x_0)$ in the limit of large
$t$ at fixed $n$.
By dimensional analysis, $\ave{s^2}(t;x_0)$ consists of a pre-factor
$(x_0^3/D)^2$ and a dimensionless function in $T=Dt/x_0^2$. 
According to the Taylor expansion \eref{s^2_expansion}, a term 
of order $T^{5/2}$ gives rise to a contribution in total linear in $x_0$,
and lower order
terms in $T$ correspond to larger powers in $x_0$. Under the provision
of large $T$ according to \Eref{s^2_expansion} the
powers are $x_0$, $x_0^3$, $x_0^4$, $x_0^5$, $x_0^6$, $x_0^7$,
$x_0^9,\ldots$. When taking the lattice Laplacian according to
\eref{lattice_Laplacian}, \Eref{s^2_expansion} is evaluated 
term by term for
$x_0\to n x_0$ and $x_0\to (n\pm1)x_0$, 
which enters in the pre-factor $(x_0^3/D)^2$ as well as in
$T(t;x_0)=tD/x_0^2$. 
Since $(n+1)-2n+(n-1)=0$, 
the first power of $x_0$ to contribute
therefore is $x_0^3$, so that
in the limit of fixed $n>0$ and large $t$
\begin{eqnarray}\fl
\ave{s_1 s_{n+1}}(t; x_0) 
\elabel{expansion_s1sn}
\\
\fl\qquad= \left(\frac{x_0^3}{D}\right)^2
\left(
  \frac{8 n}{3 \sqrt{\pi}} T^{3/2}(t;x_0)
- \frac{1+6n^2}{6} T(t;x_0)
+ \frac{n+2n^3}{3 \sqrt{\pi}} T^{1/2}(t;x_0)
+ \OC\left(T^0\right)\right) \ ,
\nonumber
\end{eqnarray}
again in terms 
of $T=T(t;x_0)=tD/x_0^2$. 
\footnote{To avoid confusion henceforth the
short hand $T$ (without arguments) is used whenever it is to be
evaluated at $t$ and $x_0$, \ie where any $n$ dependence has been
removed, $T(t;nx_0)=n^{-2} T(t;x_0)=n^{-2}T$.}
As discussed below, $\ave{s_1 s_{n+1}} - \ave{s}^2<0$ for all $T$ greater
than some threshold and $n\ne0$.

Of course, in the limit of large $n$ at fixed $t$, 
one expects asymptotic independence,
\begin{equation}
\lim_{n\to\infty} \ave{s_1 s_{n+1}}(t; x_0) =
\ave{s}^2(t; x_0) = (t x_0)^2 = \left(\frac{x_0^3}{D}\right)^2
T^2(t;x_0) \ .
\end{equation}
This large $n$ behaviour can be obtained from the small $T$ expansion,
\Eref{s2_expansion_small_T}. The polynomial part 
$T^2$
together
with the pre-factor $(x_0^3/D)^2$
contributes the expected constant $(t x_0)^2$, while $T^3(x_0^3/D)^2$ is
independent of $x_0$ and its contribution vanishes therefore after operating with the
lattice Laplacian.
The exponential
pre-factor $\exp(-1/(4T(t;n x_0)))=\exp(-n^2/(4T(t;x_0)))$ renders the
contribution from $n-1$ exponentially more relevant than from $n$ or
$n+1$:
\begin{eqnarray} \fl
\ave{s_1 s_{n+1}}(t; x_0)  =  
 (t x_0)^2 + 
\left(\frac{x_0^3}{D}\right)^2 
 \frac{256}{\sqrt{\pi}} T^{13/2} e^{-(n-1)^2\frac{1}{4T}} 
 \nonumber \\
 \times \left(
   -n^{-7} - 7 n^{-8} + 28 (2 T -1) n^{-9} + \OC\left(n^{-10}\right)
 \right)
 + \OC\left(e^{-n^2\frac{1}{4T}}\right)
 \elabel{s1sn_large_n}
\end{eqnarray}
in the limit of large $n$ at fixed $t$.

\subsubsection{Anti-correlations}
The Laplacian structure \eref{cn_laplacian} gives rise to rather
peculiar behaviour of the error of the numerical estimate of $\ave{s}$.
In a na{\"i}ve implementation one probably takes a series of $N$
samples,
$s_1,s_2,\ldots,s_N$ and estimates $\ave{s}$ by
\begin{equation}
\numave{s}=\frac{1}{N}\sum_{i=1}^N s_i \ ,
\elabel{estimator_ave_s}
\end{equation}
which is indeed an
unbiased estimator, $\ave{\numave{s}}=\ave{s}$ \cite{Brandt:98}, independent
of $N$. In the following, we will use an over-bar to indicate
numerical estimates, for example $\numave{s}$ for the estimate of $\ave{s}$.
Considering its error, for large $N$ one arrives at
a picture as cartooned in \Fref{estimate_s}: The walkers start from
$r_0(t'=0)=0, r_1(0)=x_0, r_2(0)=2x_0,\ldots,r_N(0)=Nx_0$, so that for
sufficiently large $N$ 
the very first one, $r_0$, is extremely unlikely
to interact with the very last one, $r_N$. The total area between $r_0$ and
$r_N$ is exactly the sum over all areas, 
\begin{equation}
N\numave{s}=N t x_0 + \int_0^t \rmd t'
\left[(r_N(t')-Nx_0)-r_0(t')\right] \ .
\elabel{simple_num_estimate}
\end{equation}
Assuming that the first and the
last are in fact independent, suggests that the total area is well
modelled by a Langevin type approach 
$r_{0,N}(t)-r_{0,N}(t=0)=\int_0^t \rmd t' \eta_{0,N}(t')$ 
with two independent noise sources
$\eta_{0,N}(t)$ with vanishing mean and correlator
$\ave{\eta_{0,N}(t_1)\eta_{0,N}(t_2)}=2 \Gamma^2 \delta(t_1-t_2)$. This
correlator has variance $\ave{r_{0,N}^2(t)}=2\Gamma^2t=2 D_0 t$, so
that $\Gamma^2=D_0=D/2$. Assuming $r_0$ and $r_N$ are independent, the 
variance of the numerical estimate is $2 D t^3/(3 N^2)$, which means
that the error in the estimate from the correlated sequence
$s_1,\ldots,s_N$ vanishes faster than in the fully uncorrelated case,
where the variance decays like
$\sigma^2(s)/N=32D^{1/2}t^{5/2}x_0/(15N\sqrt{\pi})+\ldots$, with
$\sigma^2(s)\equiv\ave{s^2}-\ave{s}^2$.

\begin{figure}
\begin{center}
\includegraphics[scale=0.7]{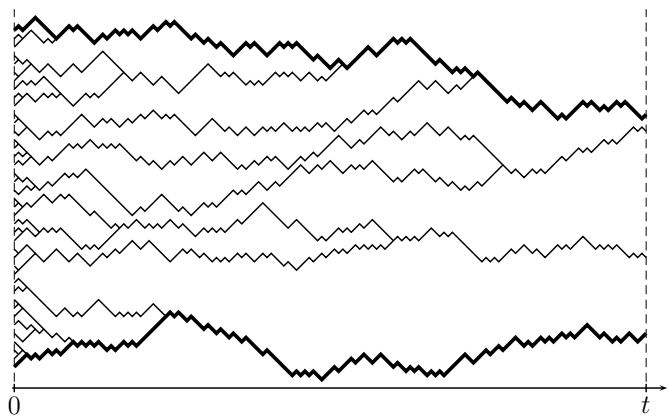}
\end{center}
\caption{
\flabel{estimate_s}
Estimating the average area $\ave{s}$ by considering a sequence
$s_1,\ldots,s_N$ is equivalent to considering the entire area enclosed
by the first and the last trajectory, $r_0$ and $r_N$ (shown in bold)
and dividing by the total number of
walkers $N$.
}
\end{figure}

This surprising behaviour is explained by anti-correlations. To see this
more clearly, we point out the ``sum rule''
\begin{eqnarray}
\fl \ave{\numave{s}^2} = 
\ave{\left(\frac{1}{N}\sum_{i=1}^N s_i \right)^2}(t;x_0) = 
\frac{2}{N^2}
\sum_{i=0}^{N-1} (N-i) \ave{s_1 s_{i+1}}(t;x_0)  - \frac{1}{N} \ave{s^2}(t;x_0)
\nonumber
\\
 \lo= 
\frac{1}{N^2} \ave{(s_1+\ldots+s_N)^2} =
\frac{1}{N^2} \ave{s^2}(t; N x_0)
\end{eqnarray}
where the equality of the first and second line is reminiscent of the
Laplacian property
\eref{cn_laplacian}. When considering large $N$, comparing
to
\eref{s2_expansion_small_T} confirms, 
\begin{equation}\elabel{estimator_variance}
\ave{\numave{s}^2} - 
\ave{\numave{s}}^2= 
\frac{2}{3N^2} D t^3 +\ldots
\end{equation}
and more importantly
\begin{eqnarray}\fl
\sum_{
i,j=1 \atop i\ne j
}^N
\left(
  \ave{s_i s_j} - \ave{s}^2 
\right)  = - N 
\sigma^2(s)(t;x_0)
+ \ave{s^2}(t; N x_0)
- N^2 \ave{s}^2(t;x_0) \nonumber 
\\ 
\lo= - N \sigma^2(s) + \frac{2}{3} D t^3 +\ldots
\end{eqnarray}
In other words: In the
variance of the estimator \eref{estimator_ave_s}, the
anti-correlations fall short of cancelling the contribution of the variance
of the area $\sigma^2(s)/N$ by only a small correction $2 Dt^3/(3N^2)$,
see \eref{estimator_variance}. 
Thus, in the limit of large $T$, the correlator $\ave{s_1
s_{n+1}}-\ave{s}^2$ drops sharply from 
\begin{equation}\fl
\sigma^2(s)(t;x_0) = 
\ave{s^2}(t;x_0) -  \ave{s}^2(t;x_0)= 
\left(\frac{x_0^3}{D}\right)^2 \left(
\frac{32}{15 \sqrt{\pi}} T^{5/2} - T^2 + \frac{8}{9 \sqrt{\pi}} T^{3/2}
+\ldots\right)
\end{equation}
at $n=0$, \Eref{s^2_expansion}, to 
\begin{equation}
\ave{s_1 s_{1+n}}(t;x_0) -  \ave{s}^2(t;x_0)= 
\left(\frac{x_0^3}{D}\right)^2 \left(
- T^2 + \frac{8}{3 \sqrt{\pi}} n T^{3/2} +\ldots\right)
\end{equation}
at fixed $n>0$, see \eref{expansion_s1sn}. Of
course, in the limit of large $n$ asymptotic independence prevails and 
only exponentially decaying
correlations are left
\begin{eqnarray}
\fl\ave{s_1 s_{1+n}}(t;x_0) -  \ave{s}^2(t;x_0)\\
\fl=
\left(\frac{x_0^3}{D}\right)^2
 \frac{256}{\sqrt{\pi}} T^{13/2} e^{-(n-1)^2\frac{1}{4T}}
 \left(
   -\frac{1}{n^7} - \frac{7}{n^8} + \frac{28 (2 T -1)}{ n^9} + \OC\left(n^{-10}\right)
 \right)
 + \OC\left(e^{-n^2\frac{1}{4T}}\right)\nonumber
\end{eqnarray}
see \eref{s1sn_large_n}. \Fref{s1sn_t12_x0eq2} shows $\ave{s_1
s_{1+n}}$ together with its numerical estimate.

\begin{figure}
\psfrag{replX}{\Large$n$}
\psfrag{replY}{\Large\hspace*{-0.5cm}$\numave{s_1s_{1+n}}/\ave{s}^2$}
\begin{center}
\includegraphics*[scale=0.55]{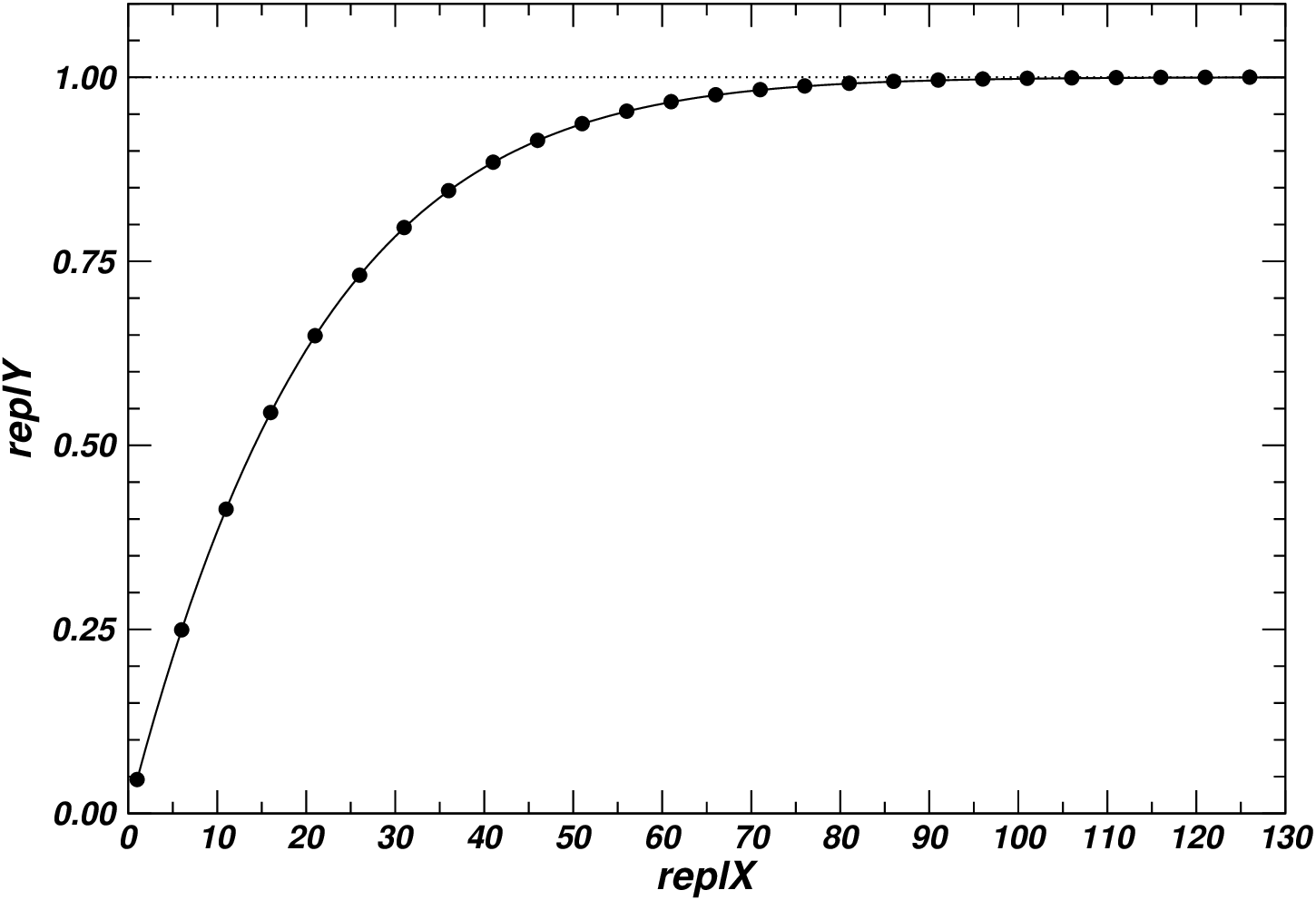}
\end{center}
\caption{\flabel{s1sn_t12_x0eq2}
The numerical estimate of the correlation function $\ave{s_1 s_{1+n}}$ for $n=1,2,\ldots,130$ for
$x_0=2$, $t=4096$ and $D=1$, shown in the form $\numave{s_1
s_{1+n}}/\ave{s}^2$ where $\ave{s}=x_0 t$. The
symbols represent numerical estimates (no error bar is shown), the straight
line is the theoretical result \eref{lattice_Laplacian}.
The second moment, $n=0$, is not shown because it is orders of magnitude
greater than $\ave{s}^2$.
}
\end{figure}

\subsubsection{Other moments}
\slabel{higher_moments}
In the pursuit to determine whether the  
correlation functions
$\ave{s_1}, \ave{s_1 s_2}, \ave{s_1 s_2 s_3},\ldots$ display gap
scaling or multiscaling in $T$, any third correlation function, other than
$\ave{s}\propto T$, \eref{first_moment}, or
$\ave{s_1s_2}\propto T^{3/2}$, \eref{expansion_s1sn}, 
is to be determined. Along the same lines
as above and using translational invariance, one finds
\begin{eqnarray}\fl
\ave{s_1 s_2 s_3}  =  \frac{1}{6} \Big(
\ave{(s_1+s_2+s_3)^3} - 2 \ave{(s_1+s_2)^3} - \ave{(s_1+s_3)^3}  + 3
\ave{s^3} \Big) \elabel{all_third} \\
\fl \qquad
=  \frac{1}{6} \Big(
\ave{(s_1+s_2+s_3)^3} - 3 \ave{s^3} 
- 6 \ave{s_1^2 s_2} - 6 \ave{s_1 s_2^2}
- 3 \ave{s_1^2 s_3} - 3 \ave{s_1 s_3^2}
\Big) \elabel{all_third_V2}
\end{eqnarray}
with the leading order of $\ave{s^3}\propto T^4$ known from
\cite{Pruessner_exactTAOM:2003,StapletonChristensen:2006}.
In case of an unbiased random walk, the second line simplifies further,
for example $\ave{s_1^2 s_2}=\ave{s_1 s_2^2}$. Unfortunately, the above
expression does not suffice to determine the leading order of $\ave{s_1
s_2 s_3}$:
On the right hand side of \Eref{all_third}
$\ave{(s_1+s_3)^3}$ is unknown, 
on the right hand side of \Eref{all_third_V2}
effectively
the same term is unknown, namely $\ave{s_1^2 s_3}+\ave{s_1 s_3^2}$. The
remaining
terms can be determined using the form suggested above,
\begin{equation}
\ave{s^3}(t;x_0) = \left(\frac{x_0^3}{D}\right)^3
\left(
C_1 T^4 + C_2 T^3 + \ldots \right) \ ,
\end{equation}
where only the leading fourth order has so far been proven to exist with
$C_1=15/8$. In particular,
\begin{equation}
\ave{s_1^2 s_2} + \ave{s_1 s_2^2}  = \frac{1}{3}\left(
\ave{(s_1+s_2)^3} - 2 \ave{s^3}
\right) = \left(\frac{x_0^3}{D}\right)^3 \left( 2 C_2 T^3 + \ldots
\right)
\end{equation}
and similarly for the leading order of
\begin{eqnarray}
\fl
6 \ave{s_1 s_2 s_3} + \ave{(s_1+s_3)^3} = 
\ave{(s_1+s_2+s_3)^3} - 2 \ave{(s_1+s_2)^3} + 3 \ave{s^3} 
\\
= 2 C_1 T^4 + 14 C_2 T^3 + \ldots \ ,
\nonumber
\end{eqnarray}
which is \eref{all_third} with all unknown terms on the left hand side
and all known terms on the right hand side.
So, if $\ave{s_1 s_2 s_3}$ has leading order higher or equal to
$\ave{(s_1+s_3)^3}$, it must be $T^4$, given that the leading
order always has positive amplitude. If it is of lower order, it might
not appear on the right hand side at all, since it might be cancelled by
a negative amplitude from $\ave{(s_1+s_3)^3}$. This is potentially a
very useful numerical criterion, because if $\ave{s_1 s_2
s_3}/\ave{(s_1+s_3)^3}$ does not converge to $0$, it means that
$\ave{s_1 s_2 s_3}$ has leading order $T^4$;
from numerics, however, it seems that the ratio asymptotically vanishes.

\subsection{Numerics}
\slabel{numerics}
All numerical simulations have been done on the lattice, all analytical
calculations in the continuum. On the lattice, in every time step 
the random walker takes
one step in the up or down direction with probabilities $p$ and
$q\equiv1-p$
respectively.\footnote{The random number generator used throughout this
study is the ``Mersenne Twister'' \cite{MatsumotoNishimura:1998b}.}
The resulting diffusion constant
is calculated from the variance after $n$ steps, $4npq=2D_0 t$ with
$t=n\Delta t$ and units chosen so that $\Delta t=1$ and
$D=2D_0=4pq$. In most of our
simulations, the walker was unbiased, $p=q=1/2$, implying $D=1$, and
the initial spacing is chosen to be minimal, $x_0=2$.\footnote{Note that
$x_i(t')$ is even at even times and odd at odd times.} This
choice could, in principle, introduce discretisation effects, which are
discussed below.

The central objective of the numerical approach was to determine the
large $T$ behaviour of $\ave{s_1 s_2 s_3}$ and $\ave{s_1 s_2 s_3 s_4}$. 
As mentioned earlier, the most na{\"i}ve implementation estimates
$\ave{s_1 s_2 \ldots s_{n+1}}$ by
\begin{equation}
\elabel{naive_est}
\numave{s_1 s_2\ldots s_{n+1}} =
\frac{1}{N-n} \sum_{i=1}^{N-n} s_{i} s_{i+1}\ldots s_{i+n}
\end{equation}
which, however, introduces correlations. To avoid that, one can produce
independent realisations at a significantly higher cost: While $N$
trajectories produce $N-n$ correlated samples, they produce only
$N/(n+1)$ uncorrelated samples, using a new set of $n+1$ trajectories
for every realisation of $s_1,s_2,\ldots,s_n$. Obviously, a correlated
sample, for instance $s_1 s_2, s_2 s_3,\ldots,s_7 s_8$ ($N=9$, $n=2$),
contains multiple uncorrelated sub-samples, for example $s_1 s_2, s_4
s_5, s_7 s_8$ but also $s_2s_3, s_5 s_6$ etc. The correlated
sample therefore converges at least as fast as its uncorrelated
sub-samples, however, possibly including anti-correlations.

It is very important to produce large samples, because of
``exponentially rare but important events'' \cite{BouchaudETAL:1990}.
For example, estimating $\ave{s_1 s_2 s_3}$ we found an
exceptionally large event where $s_1 s_2 s_3$ was almost $10^{12}$ times
bigger than the average $\ave{s_1 s_2 s_3}$ up to that point,
estimated from about $3.4 \cdot 10^{11}$ sequential samples, for $x_0 =
2$ and $t = 2^{17}$. At the end of the simulations, the inclusion of the
extreme event still increased the average $\ave{s_1 s_2 s_3}$
by a factor of $3.7$.

In addition to the statistical error, there are errors from the
discretisation and, closely related, the size of $x_0$. 
Ideally, in a numerical simulation of a continuous random walk, $x_0$
and $t$ were chosen as large as possible, which is, however, 
computationally very costly. 
To estimate the influence of lattice effects due to small $x_0$, we compared
$\numave{s_1 s_2}(t;x_0)$ to the exact result \eref{s2_exact} for
different $x_0=2,4,8,32$. The result is shown in  \Fref{s1s2} in the form
$\numave{s_1 s_2}/\ave{s_1 s_2}$ versus $t$. As expected, it suggests large $x_0$
to avoid lattice effects, which clashes with the demand
for large
$T=tD/x_0^2$ to determine large time asymptotes in $T$.
Comparing the
product of the square of the 
statistical error and the CPU-time spent on the results for different
$x_0$ indicates that a sequential, \ie correlated, see \eref{naive_est}, simulation of $x_0=2$ is most
efficient.

\begin{figure}
\psfrag{repelling}{$x_0=2$ (r)}
\psfrag{x2irw}{$x_0=2$ (i)}
\psfrag{x2srw}{$x_0=2$ (s)}
\psfrag{x4srw}{$x_0=4$ (s)}
\psfrag{x8srw}{$x_0=8$ (s)}
\psfrag{x32srw}{$x_0=32$ (s)}
\psfrag{replX}{\Large$t$}
\psfrag{replY}{\Large\hspace*{-0.5cm}$\numave{s_1s_2}/\ave{s_1s_2}$}
\begin{center}
\includegraphics*[scale=0.55]{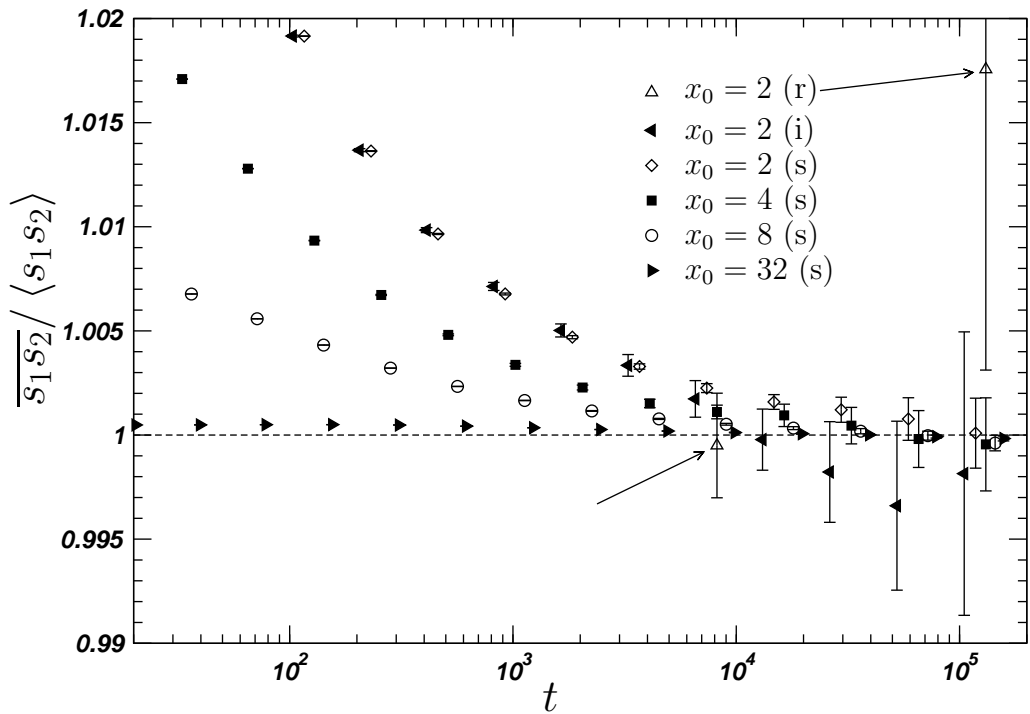}
\end{center}
\caption{\flabel{s1s2}
The ratio of the numerical estimate $\numave{s_1 s_2}$ and the
theoretical value $\ave{s_1 s_2}$. The number of samples varies between
$4\cdot 10^{10}$ and $1.16\cdot 10^{12}$.
The symbols are shifted relatively to each other to
reduce clutter, the error bars span two standard deviations in total.
The dashed line marks perfect agreement between numerics and theory.
Data marked (i) is from using independent samples, data marked with
(s) comes from sequential, correlated runs, \Eref{naive_est}. The data points marked with 
arrows have been
produced by importance sampling, (r).
}
\end{figure}

For technical reasons, the error bars shown in the graphs are estimated
from the variance of the estimator for small subsamples. 
For example, the independent 
samples were created from $1800$ individual runs, each consisting of
about
$7.2 \cdot 10^6$ to $61 \cdot 10^6$ 
independent samples totalling to $4.1 \cdot 10^{10}$ samples, 
produced by fair random walks as
described above. However, while $\ave{s_1 s_2}$ suggests satisfactory
agreement between discrete numerical and continuous analytical results,
as mentioned above, rare events made it very hard to estimate
$\ave{s_1s_2s_3}$ or higher 
correlation functions.
The most obvious way to 
overcome this problem is to use importance sampling
\cite{LandauBinder:2000} which,
however, requires a method to produce these rare, important events and a control
over their frequency. We decided to introduce a repulsive potential 
between the walkers which therefore run simultaneously rather than
sequentially. Having no analytical reference other than $\ave{s}$ and
$\ave{s_1s_2}$, it is difficult to 
assess the quality of this approach.
Given the strong impact of rare
events on $\ave{s_1 s_2 s_3}$, the quality of the numerical estimate
$\numave{s_1 s_2}$ compared to $\ave{s_1 s_2}$ has only very limited
indicative power for the quality of $\numave{s_1 s_2 s_3}$.
It is remarkable how large the error of the importance sampling scheme
is for $\numave{s_1s_2}$ compared to the na{\"i}ve methods, see
\Fref{s1s2}, and how small it is for $\numave{s_1 s_2 s_3}$, see
\Fref{s1s2s3_over_t_lin}.
Moreover, the numerically estimated error bar of any
estimator is expected to suffer from the same problems as the estimator
itself, \ie too small or wrongly biased samples. 

The key result for $\ave{s_1 s_2 s_3}$ is shown in
\Fref{s1s2s3_over_t_lin} in the form
$\ave{s_1 s_2 s_3}/[(x_0^3/D)^3 T^{3/2}]$, where we divide by
$(x_0^3/D)^3$ to render the result dimensionless and by $T^{3/2}$ to
remove the suspected leading order. It remains somewhat inconclusive whether
the fit to a logarithm in the intermediate region breaks down for the
sequential runs at
large $T$ only because of a lack of statistics. The latter is, however,
consistent with the observation that 
underestimation of $\ave{s_1 s_2 s_3}$ is more than twice as
frequent as overestimation
even when averaging over about
$4\cdot10^8$ samples.
The estimates produced from the importance sampling
method seem to confirm the presence of the logarithm even for the
largest $t=131072$. 

\begin{figure}
\psfrag{repelling}{$x_0=2$ (r)}
\psfrag{x2irw}{$x_0=2$ (i)}
\psfrag{x2srw}{$x_0=2$ (s)}
\psfrag{x4srw}{$x_0=4$ (s)}
\psfrag{x8srw}{$x_0=8$ (s)}
\psfrag{x32srw}{$x_0=32$ (s)}
\psfrag{replX}{\Large$T$}
\psfrag{replY}{\Large\hspace*{-0.5cm}$\numave{s_1s_2s_3}/[(x_0^3/D)^3
T^{3/2}]$}
\begin{center}
\includegraphics*[scale=0.55]{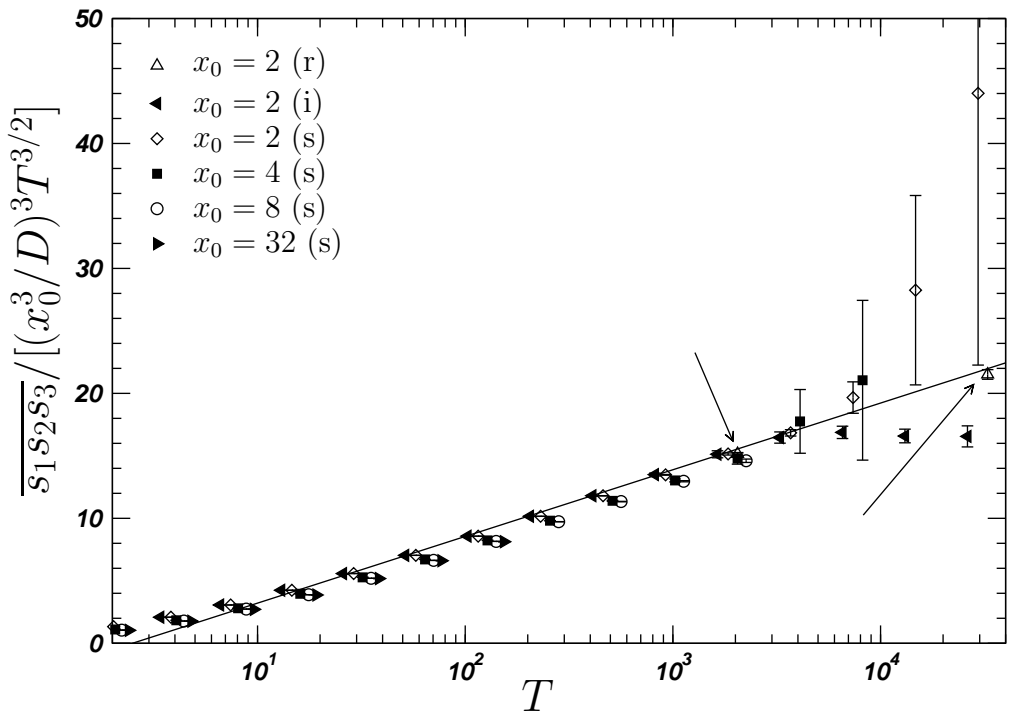}
\end{center}
\caption{\flabel{s1s2s3_over_t_lin}
The numerical estimate $\numave{s_1 s_2 s_3}$ in the form $\numave{s_1
s_2 s_3}/[(x_0^3/D)^3 T^{3/2}]$ in a log-lin plot. Symbols corresponding to different 
initial spacings $x_0$ are shifted relatively to each other to reduce
clutter. 
Data marked (i) is from independent samples, (s) from sequential samples,
(r) from importance sampling with repulsive potential.
The straight line is a fit to a
logarithm $a \ln(T/b)$ to guide the eye. The importance
sampling result (see arrows) fully agrees with the logarithmic fit in the
intermediate region of the na{\"i}ve scheme.
It remains unclear why for large $T$ the results from independent samples deviate so
significantly from the logarithmic behaviour.
}
\end{figure}

From the present result, we can quite confidently \emph{rule
out gap scaling}, which would require $\ave{s_1 s_2 s_3}/t^2$ to converge
in the limit of large $t$ to a non-vanishing value. A direct measure for
the validity of gap scaling is a moment ratio of the form
\cite{JensenPruessner:2003}
\begin{equation}
g_3 = \frac{\ave{s_1 s_2 s_3} \ave{s}}{\ave{s_1 s_2}^2}
\elabel{def_g3}
\end{equation}
which converges to a non-vanishing value if the observables obey gap
scaling, irrespective of the value of the exponents.
The numerical estimate of $g_3$ is shown in \Fref{g3} (using the
analytical results for $\ave{s}$ and $\ave{s_1 s_2}$). Its decay to $0$
indicates once more that $\ave{s_1 s_2 s_3}$ does not diverge fast
enough.

\begin{figure}
\psfrag{repelling}{$x_0=2$ (r)}
\psfrag{x2irw}{$x_0=2$ (i)}
\psfrag{x2srw}{$x_0=2$ (s)}
\psfrag{x4srw}{$x_0=4$ (s)}
\psfrag{x8srw}{$x_0=8$ (s)}
\psfrag{x32srw}{$x_0=32$ (s)}
\psfrag{replX}{\Large$T$}
\psfrag{replY}{\Large\hspace*{-0.5cm}$g_3$}
\begin{center}
\includegraphics*[scale=0.55]{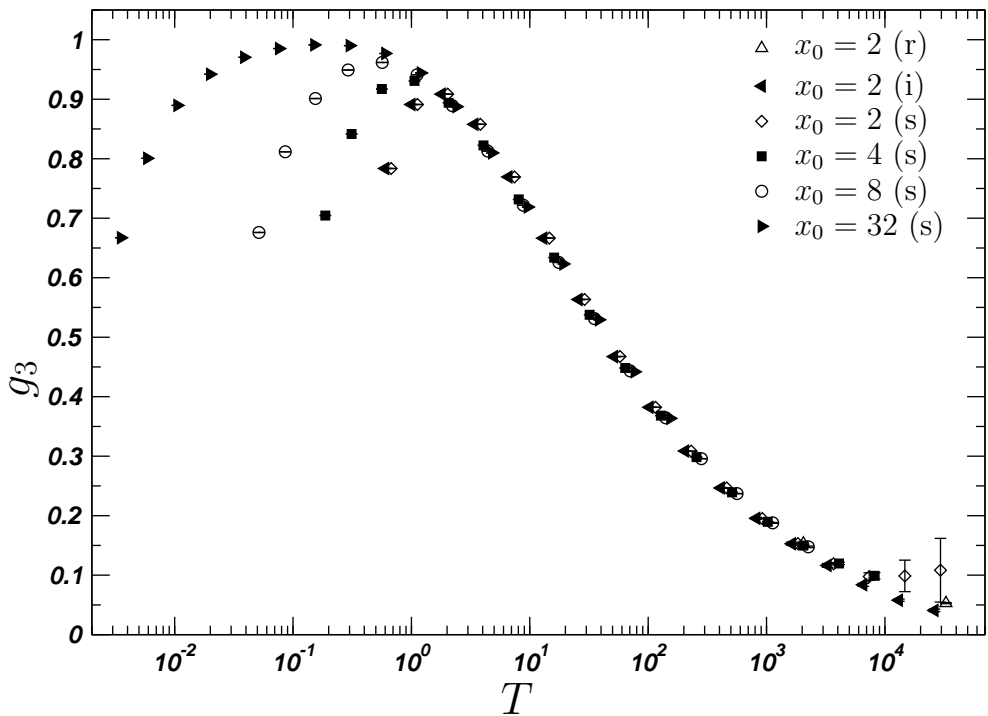}
\end{center}
\caption{\flabel{g3}
The moment ratio $g_3$, \Eref{def_g3}, would converge to a nonzero value in
the case of gap scaling. As in previous figures, data are shifted to
avoid clutter. They show different initial spacings
$x_0$ and different simulation methods [(i) independent, (s) sequential,
(r) repulsive].
}
\end{figure}

Along the same lines, \Fref{s1s2s3s4_over_t_lin} shows $\numave{s_1 s_2 s_3
s_4}/[(x_0^3/D)^4 T^{3/2}]$
together with a fit to a parabola 
in $\ln(T)$.
While the fit is not perfect and the data is, apart from the result from
importance sampling, slightly inconsistent for large $T$, the suggested
parabola seems plausible. The data is certainly not consistent with gap
scaling.

\begin{figure}
\psfrag{repelling}{$x_0=2$ (r)}
\psfrag{x2irw}{$x_0=2$ (i)}
\psfrag{x2srw}{$x_0=2$ (s)}
\psfrag{x4srw}{$x_0=4$ (s)}
\psfrag{x8srw}{$x_0=8$ (s)}
\psfrag{x32srw}{$x_0=32$ (s)}
\psfrag{replX}{\Large$T$}
\psfrag{replY}{\Large\hspace*{-0.5cm}$\numave{s_1s_2s_3s_4}/[(x_0^3/D)^4
T^{3/2}]$}
\begin{center}
\includegraphics*[scale=0.55]{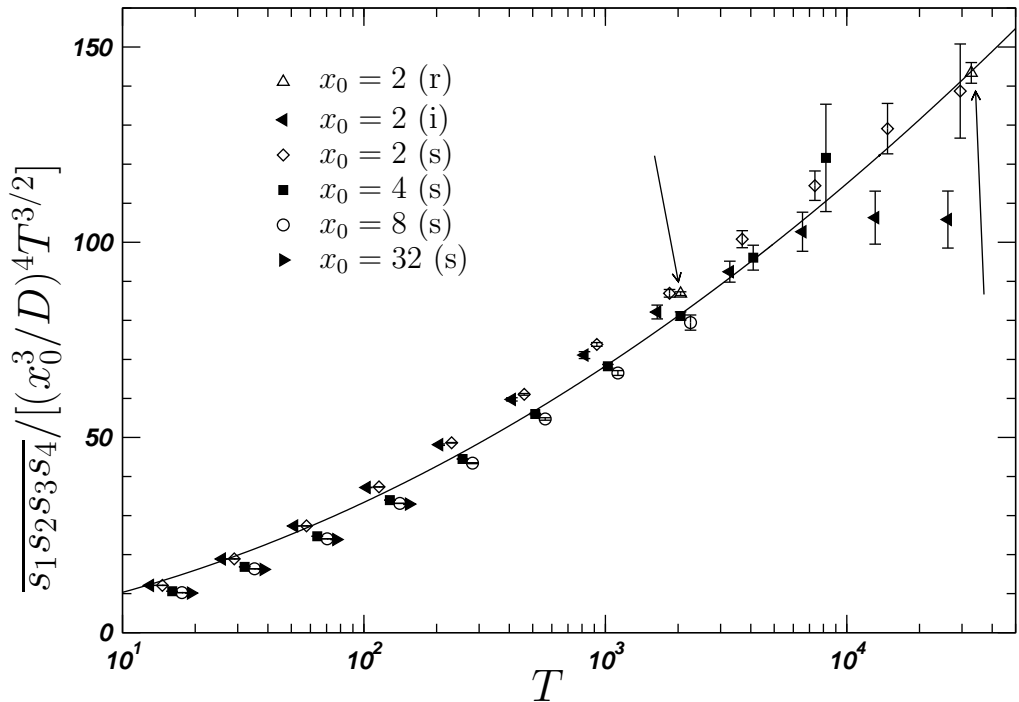}
\end{center}
\caption{\flabel{s1s2s3s4_over_t_lin}
The numerical estimate $\numave{s_1 s_2 s_3 s_4}$ in the form $\numave{s_1
s_2 s_3 s_4}/[(x_0^3/D)^4 T^{3/2}]$ in a log-lin plot,
similar to \Fref{s1s2s3_over_t_lin}. Data is shifted to avoid clutter,
and shows different initial spacings $x_0$ and different simulation
techniques.
The full line is a fit to a
parabola in the logarithm of $T$, \ie $a \ln(T/b)+ c \ln(T/b)^2$. 
}
\end{figure}

\section{Summary}
Motivated by the question whether the sequence of the areas between
trajectories of coalescing random walkers exhibits multiscaling, we have
calculated the second moment and subsequently the two point correlation function
exactly. The correlation function displays anti-correlations, as was
discovered in the analysis of the variance of the numerical estimator of
the first moment. Anti-correlations are sometimes regarded as indicating
multiscaling \cite{MunasingheETAL:2006b,MunasingheETAL:2006}.

Three moments are necessary to expose the absence of gap scaling, which
necessitated a numerical calculation. Due to the presence of rare but
important
events, convergence is slow, yet an importance sampling scheme
produces results consistent with logarithmic scaling, which is often
observed in conjunction with multiscaling
\cite{MunasingheETAL:2006b,MunasingheETAL:2006}.

While the analytical proof for multiscaling in the present system is still
lacking, the numerical results indicate it convincingly. Two possible
avenues may be pursued in the future: An exact solution along the lines
discussed in \sref{higher_moments}, or an exact calculation based on the
work by Munasinghe \etal \cite{MunasingheETAL:2006b,MunasingheETAL:2006}.

\ack{
The authors would like to thank 
A. Thomas, 
M. J. Morris,
D. Moore, N. Emarporats and R. Toumi for
technical support for the SCAN computing facility. 
GP would like to thank A. Bray and C. Connaughton for very useful
discussions, and O. Peters, R. Ecke, S. Revell and P. Watson for 
hospitality. PW would like to thank the Nuffield foundation for generous
support (Grant No. URB/33147).}

\section*{References}
\bibliography{articles,books}

\providecommand{\newblock}{}
\begin{thebibliography}{10}
\expandafter\ifx\csname url\endcsname\relax
  \def\url#1{{\tt #1}}\fi
\expandafter\ifx\csname urlprefix\endcsname\relax\def\urlprefix{URL }\fi
\providecommand{\eprint}[2][]{\url{#2}}

\bibitem{PfeutyToulouse:1977}
Pfeuty P and Toulouse G 1977 {\em Introduction to the Renormalization Group and
  to Critical Phenomena\/} (Chichester: John Wiley \& Sons)

\bibitem{Krug:1997}
Krug J 1997 {\em Adv. Phys.\/} {\bf 46} 139--282

\bibitem{benAvrahamHavlin:2000}
{ben-Avraham} D and Havlin S 2000 {\em Diffusion and Reactions in Fractals and
  Disordered Systems\/} (Cambridge, UK: Cambridge University Press)

\bibitem{Jensen:1998}
Jensen H~J 1998 {\em Self-Organized Criticality\/} (New York, NY: Cambridge
  University Press)

\bibitem{ChristensenMoloney:2005}
Christensen K and Moloney N~R 2005 {\em Complexity and Criticality\/} (London,
  UK: Imperial College Press)

\bibitem{JensenPruessner:2003}
Pruessner G and Jensen H~J 2003 {\em Phys.~Rev.~Lett.\/} {\bf 91} 244303--1--4
  (\textit{Preprint} \eprint{cond-mat/0307443})

\bibitem{Pruessner_exactTAOM:2003}
Pruessner G 2004 {\em J.~Phys.~A:~Math.~Gen.\/} {\bf 37} 7455--7471
  (\textit{Preprint} \eprint{cond-mat/0402564})

\bibitem{ConnaughtonRajeshZaboronski:2005}
Connaughton C, Rajesh R and Zaboronski O~V 2005 {\em Phys.~Rev.~Lett.\/} {\bf
  94} 194503

\bibitem{ConnaughtonRajeshZaboronski:2006}
Connaughton C, Rajesh R and Zaboronski O 2006 {\em Physica D\/} {\bf 222}
  97--115 (\textit{Preprint} \eprint{cond-mat/0510389})

\bibitem{MunasingheETAL:2006b}
Munasinghe R~M, Rajesh R and Zaboronski O~V 2006 {\em Phys.~Rev.~E\/} {\bf 73}
  051103--1--10 (\textit{Preprint} \eprint{cond-mat/0506398})

\bibitem{StapletonChristensen:2006}
Stapleton M~A and Christensen K 2006 {\em J.~Phys.~A:~Math.~Gen.\/} {\bf 39}
  9107--9126

\bibitem{MajumdarComtet:2004}
Majumdar S~N and Comtet A 2004 {\em Phys.~Rev.~Lett.\/} {\bf 92} 225501--1--4

\bibitem{MajumdarComtet:2005}
Majumdar S~N and Comtet A 2005 {\em J.~Stat.~Phys.\/} {\bf 119} 777--826

\bibitem{MunasingheETAL:2006}
Munasinghe R, Rajesh R, Tribe R and Zaboronski O 2006 {\em Commun. Math.
  Phys.\/} {\bf 268} 717--725 (\textit{Preprint} \eprint{math.PR/0512179})

\bibitem{Wegner:72}
Wegner F~J 1972 {\em Phys.~Rev.~B\/} {\bf 5} 4529--4536

\bibitem{Taeuber:2005}
T{\"a}uber U~C 2005 Critical dynamics preprint available at
  \url{http://www.phys.vt.edu/~tauber/utaeuber.html}

\bibitem{Brandt:98}
Brandt S 1998 {\em Data Analysis\/} (Berlin Heidelberg New York:
  Springer-Verlag)

\bibitem{MatsumotoNishimura:1998b}
Matsumoto M and Nishimura T 1998 {\em {M}onte {C}arlo and Quasi-{M}onte {C}arlo
  Methods 1998\/} (Berlin Heidelberg New York: Springer-Verlag) preprint from
  {\tt http://www.math.h.kyoto-u.ac.jp/\~\ $\!\!\!$matumoto/RAND/DC/dc.html}
  \urlprefix\url{http://www.math.h.kyoto-u.ac.jp/~matumoto/RAND/DC/dc.html}

\bibitem{BouchaudETAL:1990}
Bouchaud J~P, Comtet A, Georges A and {Le Doussal} P 1990 {\em Ann. Phys.\/}
  283--341

\bibitem{LandauBinder:2000}
Landau D~P and Binder K 2000 {\em A Guide to {M}onte {C}arlo Simulations in
  Statistical Physics\/} (Cambridge, UK: Cambridge University Press)

\end{thebibliography}

\end{document}